\documentclass[12pt,epsfig]{article}
\usepackage{amsmath}
\usepackage{amssymb}
\usepackage{graphicx}
\usepackage{epsfig}
\newcommand{\bea}{\begin{eqnarray}}

\newcommand{\eea}{\end{eqnarray}}

\newcommand{\ep}{\epsilon}

\def\bes{\begin{eqnarray}}
 \def\ees{\end{eqnarray}}
\def\be{\begin{equation}}
\def\ee{\end{equation}}
\def\bs{\begin{subequations}}
\def\es{\end{subequations}}
\newcommand{\een}{\end{subequations}}
\newcommand{\ben}{\begin{subequations}}
\newcommand{\beq}{\begin{eqalignno}}
\newcommand{\eeq}{\end{eqalignno}}

\def\Gin{{\left(G^{-1}_0\right)}}
\def\trr{{\rm tr}}

 \def\picl{\pi_{cl}}
 \def\dpi{{\delta\pi}}
 \def\dG{{\delta G}}

 \def\ex{\epsilon}
 
 \def\Lx{\Lambda}

\newcommand{\calk}{{\cal K}}

\begin{document}


\begin{center}
{ \Large \bf
Quantum Corrections in Classicalon Theories}
\\
\vspace{1.5cm}
{\Large 
P. Asimakis, N. Brouzakis, A. Katsis and N. Tetradis 
} 
\\
\vspace{0.5cm}
{\it
Department of Physics, University of Athens, Zographou 157 84, Greece
} 
\end{center}
\vspace{3cm}
\abstract{
We use the heat kernel in order to compute the one-loop effective action on a classicalon background.
We find that the UV divergences
are suppressed relative to the predictions of standard perturbation theory in the interior of the classicalon. 
There is a strong analogy with the suppression of quantum fluctuations in Galileon theories, within the regions where the
Vainshtein mechanism operates (discussed in arXiv:1401.2775). Both classicalon and Galileon theories display 
reduced UV sensitivity on certain backgrounds. 
}

\newpage


The scenario of classicalization \cite{dvali} suggests that
high-energy scattering in certain classes of nonrenormalizable scalar field theories can take place at length scales much larger than the
typical scale associated with the nonrenormalizable terms in the Lagrangian.
It has been argued that the reason for this behavior is that the UV completion of the theory is achieved not through the
inclusion of arbitrarily hard modes, but through collective states, which are composed of a large number of soft 
quanta and display classical properties \cite{dvali2}. 
The crucial ingredient is the presence of a semiclassical configuration, the classicalon, generated by a point-like source.
Classicalons generally exist in theories of Goldstone bosons, or other higher-derivative theories \cite{dvali,nonlinear}.
In spite of several studies, a complete picture of classicalization is not available yet.
It has been shown that a collapsing spherical wavepacket can be deformed significantly
at the so-called classicalization radius, which can be much larger than the fundamental length scale of the theory \cite{dvpirts,rizos}.
However, in some theories the classical scattering problem may not have
real solutions over the whole space at late times\footnote{An interesting possibility 
is that the absence of a real classical solution in the scattering problem may indicate
the presence of a tunnelling solution in the quantum theory \cite{tsolias}, so that classicalization is a quantum process.}, 
while in others the maximum of the collapsing wavepacket
can reach distances of the order of the fundamental scale. It seems that classicalization is not a generic phenomenon, but
appears in theories with particular properties. It has been suggested that such theories cannot be extended 
through the inclusion of new degrees of freedom at short scales, but generate a physical 
UV cutoff through their own dynamics \cite{dfg}. The ``wrong-sign" DBI theory is a possible candidate. 
It can display some undesirable features, such as
superluminality on certain nontrivial backgrounds \cite{inflation1,adams} (see, however, \cite{dfg,babichev}). 
On the other hand, it has been argued that quantum fluctuations
are suppressed in  this theory, as well as in all theories that admit classicalons \cite{vikman}. This is consistent with the
notion that hard modes do not play a role in high-energy scattering. In this letter we would like to 
address this issue through an explicit calculation of quantum corrections on classicalon backgrounds.

We shall follow the steps of a similar calculation, performed in the context of the cubic Galileon theory on a background that realizes the
Vainshtein mechanism \cite{vainshtein}.
The Galileon theory describes the dynamics of the scalar mode that 
survives in the decoupling limit of the DGP model \cite{dgp}. It contains 
a dimensionful coupling that sets the scale $\Lx$ at which the theory becomes strongly coupled \cite{galileon}. 
This scale can be identified with the UV cutoff. In the presence of a point-like source,
the theory has a spherically symmetric solution with a characteristic radius $r_V$, usually refer to as the
Vainshtein radius \cite{vainshtein}. At distances much larger than $r_V$ classical
fluctuations on top of the background propagate as free waves, while at distances smaller than $r_V$ they are suppressed.
In \cite{quantum1} it was argued that, at the scales at which the Vainshtein mechanism operates, quantum fluctuations 
could be suppressed as well. 
Quantum corrections in Galileon theories were studied in refs. \cite{ctzb,suppr} on a trivial background.
In \cite{suppr} the one-loop corrections were calculated  on the Vainshtein background in the presence of an explicit UV cutoff.  
Through an appropriate modification of the heat-kernel formalism, it was shown that 
the background reduces the magnitude of the divergent terms. It must be emphasized that the theory remains nonrenormalizable.
However, the sensitivity to the physical UV cutoff  is much smaller than what would have been expected through naive 
perturbative arguments.

The similar features of Galileon and classicalon theories make it plausible that a mechanism of suppression of quantum fluctuations 
could operate on classicalon backgrounds. In order to examine this possibility, 
we repeat the calculation of ref. \cite{suppr} for theories that can support classicalons. We consider a class of 
actions of the form   
\be
S=\int d^4x \, {\cal K}  \left( X \right),
\label{genaction}
\ee
with $X=\partial_{\mu} \pi \partial^{\mu} \pi /2$.
Our convention for the Minkowski metric is $\eta_{\mu\nu}={\rm diag}(-1,1,1,1)$.
The equation of motion has a one-parameter, static, spherically-symmetric solution given by
\be
\calk_X\pi'(w)=-\frac{c}{2w^{\frac{3}{2}}},
\label{solclas} \ee
with  $c$ an integration constant (positive or negative), 
$w=r^2$, $X=2w \pi'^2$, and $\calk_{X}= \calk'\left(X\right), \, \calk_{XX}=\calk'' \left(X\right)$ etc.
The primes indicate derivatives with respect to the indicated arguments of the various functions: $\pi'(w)=d\pi/dw$,
$\calk'(X)=d\calk/dX$. The factor of 2 in the denominator and the minus sign have been added in order to simplify formulae in the following.
When these configurations extend over the whole space, they can be identified as classicalons.

For specific calculations we concentrate on variations of the DBI action. 
The standard DBI action has 
\be 
\calk_1=\frac{1}{\mu} \sqrt{1-2\mu X}
\label{dbichi} \ee
 with $\mu<0$, while the ``wrong-sign" theory corresponds to
$\mu>0$.  The solution (\ref{solclas}) becomes
\be
\pi_1'(w)= \frac{1}{2} \frac{c}{\sqrt{w^3+\mu c^2 w}},
\label{soldbi} \ee
with $c$ positive or negative.
For $\mu<0$ the two branches can be joined at the location of the square-root singularity in order to obtain the 
catenoidal solution that has been studied in \cite{gibbons}. This solution does not extend over the whole space and it
is not possible to characterize it as a classicalon. On the other hand, the ``wrong-sign" DBI theory with $\mu>0$ leads to configurations that
span the whole space. These are the classicalons considered in \cite{dvali}. 
The discontinuity of the first derivative at the origin requires the presence of a $\delta$-function source 
at this point. 
Similar solutions can be obtained for a theory with 
\be
\calk_2=-X-\mu X^2/2
\label{quadchi} \ee
and $\mu>0$
 (keeping only the first two terms in the expansion of the
square root in the DBI theory). They are given by
\be
\pi_2'(w)=\frac{2\times 3^{\frac{1}{3}}\,\mu w^4-\left(   - 9\mu^2 c w^5+\sqrt{\mu^3 w^{10}\left(24w^2+81\mu c^2 \right)}  
 \right)^{\frac{2}{3}}}{2\times 3^{\frac{2}{3}}\,\mu  w^{\frac{5}{2}}
\left(       - 9\mu^2 c w^5+\sqrt{\mu^3 w^{10}\left(24w^2+81\mu c^2 \right)}                 \right)^{\frac{1}{3}}}.  
\label{solquart} \ee

\begin{figure}[t]
\begin{center}
 \includegraphics[width=14.0cm]{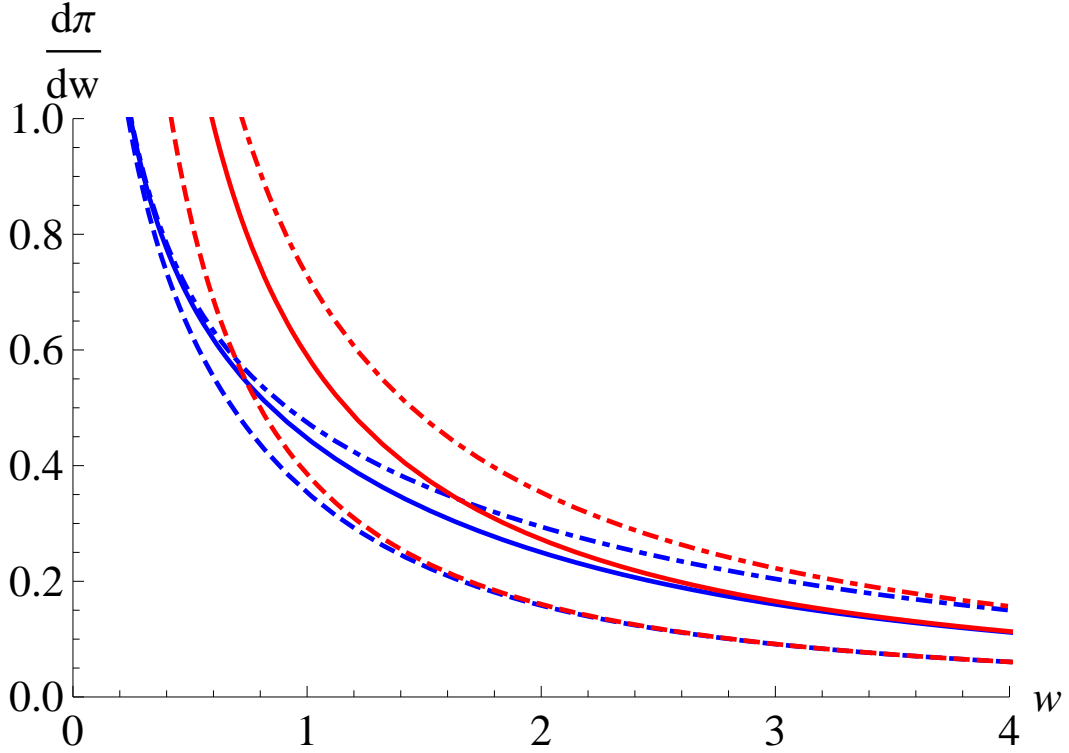}%
\end{center}
\caption{The classicalon solutions $\pi_1'(w)$ (blue lines) and $\pi_2'(w)$ (red lines) for $\mu=1$ and 
$c=1$ (dashed lines), $c=2$ (solid lines), $c=3$ (dot-dashed lines). }
 \label{figg}
 \end{figure}

In fig. \ref{figg} we depict the classicalon solutions $\pi_1'(w)$ (blue lines) and $\pi_2'(w)$ (red lines) for 
various values of the integration constant $c$. We express all dimensionful quantities in terms of the fundamental scale of the
theory, so that $\mu=1$. 
For large $w$, both solutions are approximately given by $\pi'(w)\simeq c/(2w^{3/2})$, so that  $\pi(r)=-c/r$.
On the other hand, for $\mu>0$ and small $w$ we have $\pi_1'(w)\simeq {\rm sign}(c)/(2\sqrt{\mu w})$ and 
 $\pi_2'(w)\simeq {\rm sign}(c)\,|c|^{1/3}/(2^{2/3}\mu^{1/3} w^{5/6})$.
The transition between the two regimes occurs at the classicalization radius $r_{cl}=\sqrt{w_{cl}}\sim (c^2\mu)^{1/4}$. 
We do not consider the structure of the classicalons at distances from the origin
smaller than $\sim \mu^{1/4}$ because we assume that the theory contains a physical UV cutoff $\Lambda \sim \mu^{-1/4}$.
For $|c|\gg 1$ there is a hierarchy between the scales $\mu^{1/4}$ and $r_{cl}$, and the classicalons are well defined classical objects.

Our aim is to evaluate the one-loop effective action 
\be
\Gamma_{1}= \frac{1}{2} {\rm tr} \log \Delta_E,
\label{gamma1} \ee
where $\Delta_E$ is the fluctuation operator on the classicalon configuration. The 
calculation of the effective action (\ref{gamma1}) requires the transition to Euclidean signature though the
definition $t=-ix^0$. For this reason the derivative operators appearing in $\Delta_E$ are assumed to act on fields
in four-dimensional Euclidean space.
The second variation of the action (\ref{genaction}) around the solution (\ref{solclas}) gives
\begin{eqnarray} 
\Delta_E&=&-G_{\mu \nu} \,  \partial^{\mu} \partial ^{\nu}-E_\mu \, \partial^\mu
\label{flucop1} \\
G_{\mu\nu}&=&-\calk_{X}\,  g_{\mu\nu} -\calk_{XX} \, \partial_{\mu} \pi \, \partial{_\nu} \pi \,
\label{flucop2} \\
E_\mu&=&-2 \calk_{XX}\, \partial_{\mu}\partial_\nu \pi\, \partial ^{\nu}\pi
- \calk_{XXX} \, \partial_\nu \partial _{\rho}\pi\,  \partial^{\rho} \pi\,  \partial^{\nu} \pi \, \partial _{\mu}\pi
 - \calk_{XX}\, \Box \pi\, \partial_{\mu} \pi  .
\label{flucop} \end{eqnarray}
Here $g_{\mu\nu}$ stands for the Euclidean metric.

We would like to compute the effective action (\ref{gamma1}) using the heat kernel \cite{vassilevich}.
The calculation of $\trr \log \Delta_E$ for the fluctuation operator  
(\ref{flucop1}) can be mapped onto the calculation for a similar operator with 
covariant derivatives involving both a Riemann and a gauge part \cite{vassilevich}, for which known results exist \cite{salcedo}.
However, the correspondence between the two pictures is very complicated. We find it more efficient to 
follow the approach of \cite{nepomechie}, as applied to the case of Galileon theories in \cite{suppr}.
The heat kernel of $\Delta_E$ can be computed through the relation
\be 
h(x,x',\ep)= \int \frac{d^4k}{(2 \pi)^4} e^{-ikx'}e^{-\ep \Delta_E} e^{ikx}.
\label{heatk} \ee
The effective action can then be obtained from its diagonal part as
\be
\Gamma_1=-\frac{1}{2}\int_{1/\Lx^2}^{\infty} \frac{d \ex}{\ex} \int d^4x\,  h(x,x,\ex).
\label{aaa} \ee
A lower limit has been introduced for the $\ex$-integration in order to regulate the possible UV divergences. In our case, 
the UV cutoff is assumed to be $\Lx\sim \mu^{-1/4}$.
The divergent terms in the effective action are generated through the expansion of the exponential in eq. (\ref{heatk}).
In order to determine the UV divergences, which appear for $\ex\to 0$, 
it is useful to rescale $k$ by $\sqrt{\ex}$, as was done in ref. \cite{nepomechie}. (For details, see \cite{suppr}.)
The diagonal part of the heat kernel becomes
\be
h(x,x,\ep)=\int\frac{d^4k}{(2 \pi)^4}\frac{1}{\ex^2}\exp \left\{
 - G_{\mu\nu}k^\mu k^\nu+
2i\sqrt{\ex} G_{\mu \nu} k^\mu\partial^\nu+i\sqrt{\ex} E_\mu k^\mu 
+\ex G_{\mu \nu} \partial^\mu\partial^\nu + \ex E_\mu \partial^\mu
 \right\},
\label{diagongen}
\ee
where it is assumed that it acts on a function $f(x)=1$.
The momenta can be transformed as $k^\mu=S^\mu_{~\nu} k'^\nu$,
with $S$ satisfying
$S^T G S=g$, where $g$ stands for the Euclidean metric.
The first term in the exponent now takes the simple form $-k'^2$. It is not possible, however, 
to isolate immediately a term $\exp(-k'^2)$ because
$k'^\mu$ does not commute with the
derivative operators. The
Baker-Campbell-Hausdorff formula must be employed, as explained in \cite{suppr}. Finally, the effect of a nontrivial background can be
studied by writing the field as $\pi=\picl+\dpi$ and expanding in powers of $\dpi$.

The divergent terms in the effective action are generated through an expansion in powers of $\ex$. 
In this letter we consider only the leading order, as the structure of the subleading terms is very 
complicated. However, the leading contribution is sufficient to demonstrate the effect of the background on the 
divergent terms. A discussion of all such terms for the cubic Galileon theory can be found in \cite{suppr}.
At order $\ex^{-2}$ the diagonal part of the heat kernel is simply
\be
h(x,x,\ex)=\frac{1}{16\pi^2}\frac{1}{\ex^2} \det S=\frac{1}{16\pi^2}\frac{1}{\ex^2} (\det G)^{-\frac{1}{2}}.
\label{heatgen} \ee
We can split the field as $\pi=\picl+\dpi$, in terms of the background $\picl$ and small fluctuations around it. Then the 
determinant can be written as 
\be
\det[G_0+\dG]=\det G_0 \left( 
1+{\rm tr}[G_0^{-1}\, \dG]-\frac{1}{2} {\rm tr}[(G_0^{-1}\, \dG)^2]+\frac{1}{2} \left({\rm tr}[G_0^{-1}\, \dG]\right)^2 + ...
\right),
\label{expdet} \ee
with $G_0=G(\picl)$ and $\dG$ expanded in powers of $\dpi$.

For theories with actions of the form (\ref{genaction}) and a spherically symmetric background $\picl$ we have
\be
G_0={\rm diag} \left(-\calk_X,-\calk_X,-\calk_X,-\calk_X-4w \calk_{XX} \pi'^2 \right).
\label{genmet0} \ee
We use a Cartesian system of coordinates with its origin at the center of the classicalon.
Without loss of generality, we evaluate the determinant at a point along the $z$-axis.
The classicalon background (\ref{solclas}) allows us to express $\calk_X$ and $\calk_{XX}$ in terms of $\picl$. In this way we
obtain
\begin{eqnarray}
G_0&=& \frac{c}{2w^{\frac{3}{2}}\picl'}\,  {\rm diag} \left(1,1,1, 1-\frac{3\picl'+2w\picl''}{\picl'+2w\picl''}\right)
\label{G0} \\
\left( {\det}G_0 \right)^{-\frac{1}{2}}&=&\left( \frac{2w^{\frac{3}{2}}\picl'}{c} \right)^2\left(1-\frac{3\picl'+2w\picl''}{\picl'+2w\picl''} \right)^{-\frac{1}{2}}.
\label{detG0} \end{eqnarray}
Using eq. (\ref{heatgen}) in order to perform the $\ex$-integration in eq. (\ref{aaa})  reproduces the quartic divergence of the 
vacuum energy density $\sim \Lx^4 $. The determinant factor (\ref{detG0}) introduces a radial dependence for this density. 
At large distances from the center of the classicalon, the higher-derivative terms in the action are expected to become 
subleading to the standard kinetic term, so that $\calk_X\simeq 1$. It is apparent from eq. (\ref{solclas}) that 
the field obeys $\picl'\simeq c/(2w^{3/2})$. This expectation is realized by the explicit solutions of eqs. (\ref{soldbi}) and (\ref{solquart}).
Therefore, for large $w$ we obtain ${\det} G_0\simeq 1$, and the standard result for the vacuum energy density is reproduced.
On the other hand, for small $w$ the first factor in the rhs of eq. (\ref{detG0}) is expected to become small and suppress the 
quantum contribution to the vacuum energy. The exact $w$-dependence of ${\det} G_0$ is model dependent, as both factors in the
rhs of eq. (\ref{detG0}) may play a role. For the ``wrong"-sign DBI theory of eq. (\ref{dbichi}) with $\mu>0$ we find 
\be
\left( {\det}G_0 \right)^{-\frac{1}{2}}=\frac{w^3}{(w^2+c^2\mu)^{\frac{3}{2}}}.
\label{detG0dbi} \ee
The suppression at small $w$ is very strong: $\left( {\det}G_0 \right)^{-{1}/{2}}\sim (w/w_{cl})^3=(r/r_{cl})^6$, where we 
have defined the classicalization radius $w_{cl}=\sqrt{c^2\mu}$. In the case of the theory of eq. (\ref{quadchi}) the 
analytical expressions are more complicated and we do not display them. 
The suppression factor is $\left( {\det}G_0 \right)^{-{1}/{2}}\sim (w/w_{cl})^{4/3}=(r/r_{cl})^{8/3}$. 

The field-dependent part of the effective action is also suppressed by the background. The complete analysis is complicated, while
the final expressions are model dependent and not particularly illuminating. For the Galileon theory on a Vainshtein background this analysis 
has been performed in \cite{suppr} for the terms quadratic in the fluctuation $\dpi$ around the background.
In order to demonstrate the suppression mechanism in theories that contain classicalons, we examine some of the quartically
divergent corrections for the theory of eq. (\ref{quadchi}). We have
\be
\delta G_{\mu\nu}\simeq\mu\left(\partial_\rho \picl \partial^\rho \dpi \, g_{\mu\nu}+\partial_\mu \picl \partial_\nu \dpi +
\partial_\nu \picl \partial_\mu \dpi  \right)
+\mu \left(
\frac{1}{2}\partial_\rho\dpi \partial^\rho \dpi \, g_{\mu\nu}+\partial_\mu\dpi \partial_\nu \dpi \right)+{\cal O}\left( \dpi^3\right),
\label{deltG} \ee
with $g_{\mu\nu}$ the Euclidean metric.
For a diagonal $G_{\mu\nu}$ of the form (\ref{genmet0}), we find 
\begin{eqnarray}
&&\det[G_0+\dG]=\det G_0 \Biggl[ 
1+\mu\Gin_{\mu\nu}\partial^\mu\dpi\partial^\nu\dpi+\frac{1}{2}\mu\,\trr\Gin\, \partial_\mu\dpi\partial^\mu\dpi
\nonumber \\
&&-4\mu^2 w \picl'^2 \Gin_{33}\Gin_{\mu\nu}\, \partial^\mu\dpi\partial^\nu\dpi
\nonumber \\
&&+2\mu^2w\picl'^2  \left[  \left[  \trr  \Gin \right]^2  -\trr \left[ \Gin^2\right] +4\,\trr\Gin \Gin_{33}-2\,\left[\Gin_{33} \right]^2
\right) \, \partial_3\dpi\partial^3\dpi
\Biggr].
\nonumber \\
&~&
\label{detexp} \end{eqnarray}
As before, we use a Cartesian system of coordinates whose origin lies 
at the center of the classicalon, and evaluate the determinant at a point along the $z$-direction.
The Euclidean symmetry of the action of fluctuations is broken by the classicalon background.
For small $w$ the various element of $G^{-1}_0$, as well as its trace, scale as $\left[G^{-1}_0\right]\sim w^{2/3}$. We also have
$\left[\det G_0\right]\sim w^{-8/3}$ and $\left[ w\picl'^2\right]\sim w^{-2/3}$.
When expanding $\left(\det[G_0+\dG]\right)^{-1/2}$ in powers of $\dpi$, we obtain an overall suppression factor $(\det G_0)^{-1/2}\sim w^{4/3}$.
Moreover, all terms of order $\dpi^2$ receive an additional suppression $\sim w^{2/3}$. We point out that the scale below which the suppression
takes place is set by the classicalization radius $r_{cl}=\sqrt{w_{cl}}\sim |c|\mu^{1/4}\gg \mu^{1/4}$.

An exhaustive analysis of all the terms in the effective action is beyond the scope of this letter. 
Our main aim here has been to demonstrate the similarity of the structure of quantum corrections in classicalon and Galileon theories. 
We emphasize that the theories remain nonrenormalizable in the technical sense, as the cutoff dependence is not eliminated. For this reason, the analysis we
presented is meaningful only if the UV cutoff is physical. In such a case, a nontrivial background can induce a significant suppression of
the quantum corrections. This behavior does not occur in renormalizable theories, in which the background has an effect only on the
IR regime. Our findings provide support to the claim that the UV sensitivity of the special theories that contain classicalons can be reduced, if these
configurations dominate the high-energy processes. The quantum corrections are suppressed in the interior of the classicalon, so that they
do not lead to substantial modifications of the background.
However, the complete inability to probe the UV regime remains a speculation that
requires further analysis within a more realistic theory along the lines we outlined in this work.

\section*{Acknowledgments}
The work of N.B. and N.T. has been co-financed by the European Union (European Social Fund – ESF) and Greek national 
funds through the Operational Program ``Education and Lifelong Learning" of the National Strategic Reference 
Framework (NSRF) - Research Funding Program: ``THALIS. Investing in the society of knowledge through the 
European Social Fund".

\end{document}